\documentclass{aa}
\usepackage[latin1]{inputenc}
\usepackage{graphics}

\makeatletter

\providecommand{\LyX}{L\kern-.1667em\lower.25em\hbox{Y}\kern-.125emX\@}

\usepackage{aabib99}

\makeatother

\begin{document}

\thesaurus{06 (03.09.2; 03.13.5; 03.13.7;03.20.2)}

\title{First images on the sky from a hyper telescope\thanks{
Based on observations performed at the Observatoire de Haute Provence
}}

\author{E. Pedretti \inst{1,2} \and
          A. Labeyrie \inst{1,2} \and       
          L. Arnold \inst{2} \and       
          N. Thureau \inst{3} \and       
          O. Lardiere \inst{2} \and       
          A. Boccaletti \inst{1,4} \and
          P. Riaud \inst{2,4} }

\offprints{E. Pedretti}

\institute{Coll\`{e}ge de France, Paris11 place Marcelin Berthelot, F-75231 Paris Cedex
05\and
      Observatoire de Haute Provence, CNRS, F-04870, Saint Michel l'Observatoire\and
      Observatoire de la C\^{o}te d'Azur, D\'epartment Fresnel, ISA-GI2T, F-06108
Saint Vallier de Thiey\and
      DESPA, Observatoire de Paris-Meudon, 5 Pl. J. Janssen, F-92195 Meudon}

\mail{pedretti@obs-hp.fr}

\date{Received April 27; Accepted ?}

\maketitle
\begin{abstract}
We show star \textbf{}images obtained with \textbf{}a \textbf{}miniature \textbf{}``densified
pupil imaging interferometer'' also called a hyper-telescope. The formation
of such images violates a ``golden rule of imaging interferometers'' \textbf{}which
appeared to forbid the use of interferometric arrangements differing from a
Fizeau interferometer. These produce useless images when the sub-apertures spacing
is much wider than their size, owing to diffraction through the sub-apertures.
The hyper-telescope arrangement solves these problems opening the way towards
multi-kilometer imaging arrays in space.  We experimentally obtain an intensity
gain of \( 24\pm 3\times  \) when a densified-pupil interferometer is compared
to an equivalent Fizeau-type interferometer and show images of the double star
\textbf{}\( \alpha  \)~Gem. The initial results presented confirm the possibility
of directly obtaining high resolution and high dynamic range images in the recombined
focal plane of a large interferometer if enough elements are used.

\end{abstract}

\section{Introduction}

The possibility of obtaining direct usable images at the combined focus of a
diluted interferometric array was long overlooked. Theoretical work on densified
pupil interferometry \cite{Labeyrieb,Labeyriec}, numerical \textbf{}simulations
of its imaging properties \cite{Boccaletti} and a new algorithm for co-phasing
a diluted array of several apertures \cite{Pedrettia,Pedrettib} now open a
new \textbf{}evolutionary track towards large optical arrays in space and on
ground. 

Fizeau optical arrays, the equivalent of a telescope carrying a multi-hole aperture
mask, have imaging properties similar to telescopes. If Rayleigh's criterion
is met, there is a peaked spread function which becomes convolved with features
of the observed object. Field is infinite in principle, although limited by
telescope aberrations. If the holes in the mask are much smaller than their
spacing however, the spread function has a vast halo of diffracted light surrounding
its central interference peak. This removes most energy from the peak and creates
a useless continuous level in the image. The ensuing image degradation would
become disastrous in the giant systems, spreading across kilometers, considered
for space interferometry with metre-sized mirrors as aperture elements
\begin{figure}
{\par\centering \resizebox*{1\columnwidth}{!}{\includegraphics{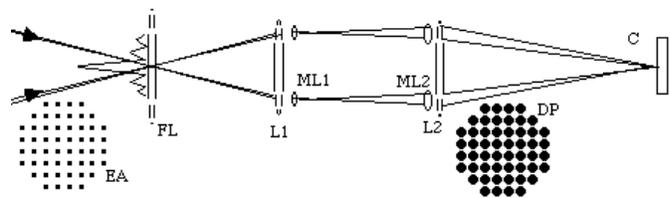}} \par}

\caption{\label{fig: densifier}Pupil densification with micro lenses. The \textbf{}converging
beams coming from widely separated sub\textbf{-}apertures (EA) are collimated
by lens L1. An image of the entrance aperture is formed, by field lens \textbf{}FL\textbf{,}
on the micro-lenses array \textbf{}ML1 having a shorter focal length than \textbf{}the
\textbf{}second micro\textbf{-}lens array ML2 located downstream. The focal
length ratio of \textbf{}ML2 and ML1 is the pupil densification factor. \textbf{}With
suitable phasing adjustments L2 forms a directly exploitable image on the detector
C (\copyright PASP).}
\end{figure}

Michelson \cite{Michelson_21}, in his 20-feet beam, avoided the halo problem
by densifying the exit pupil with his four-mirror periscopic arrangement, i.e,
giving it a higher sub-pupil size/spacing ratio than in the entrance aperture.
This caused the spread function to lose its field invariance: fringes were moving
across the diffractive halo if a point source moved. The convolution thus no
longer applied. Extended versions of Michelson's beam, using many apertures
instead of two, were considered since the 1970's, but the loss of the convolution
relation appeared to make them incapable of forming direct images \textbf{}\cite{Labeyriea,Beckers}.
The point was formalized into a ``golden rule of imaging interferometry''
\cite{Traub_86}. 

\begin{figure*}
{\par\centering \resizebox*{0.9\textwidth}{!}{\includegraphics{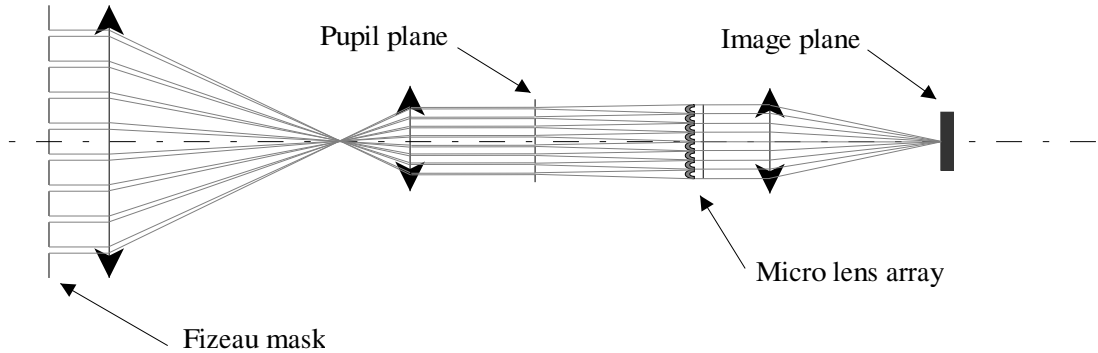}} \par}

\caption{\label{fig: opt}Miniature \textbf{}hyper-telescope. A Fizeau mask is placed
in front of an afocal diffracting telescope. Its image in the exit pupil has
narrow sub-pupils which diffract light so that facing elements of the micro-lens
array located downstream \textbf{}are each filled by the corresponding diffracted
lobe. The pupil-array distance matches the focal distance of the micro-lenses
in the array, so that the diverging diffracted beams are approximately re-collimated,
\textbf{}thus achieving the desired pupil densification\textbf{.}}
\end{figure*}
Only much later did it appear that the rule can be evaded \cite{Labeyrieb},
with considerable benefit in terms of future applications. It was shown that
a densified pupil such as Michelson's arrangement, but incorporating many elements,
can provide direct images if the sub-pupil centres are preserved in terms of
their relative locations. The image formation may then be described as a ``pseudo-convolution''
\cite{Labeyrieb}, \textbf{}where two functions:

\begin{itemize}
\item a field-invariant \char`\"{}interference function\char`\"{}, which is the Fourier
transform of the sub-pupil centres array,
\item a broad \char`\"{}diffraction function\char`\"{} which is the Fourier transform
of a single sub-pupil,
\end{itemize}
move at different speed when a point source moves away from the central axis
of the instrument. The interference function moves faster than the diffraction
function, the speed depending on the densification factor. If the densification
factor is large then the movement of the diffraction function with respect to
the interference function, becomes negligible\footnote{%
If the densification factor is one, the two \textbf{}functions move at the same
speed and the pseudoconvolution becomes a normal convolution. This is the case
of the Fizeau interferometer.
} .

The diffractive envelope can then be considered a static windowing function,
limiting the angular span of the interference function. Inside the envelope,
an ordinary convolution of the object with the interference function, takes
place \textbf{}. 

Systems other than the Michelson beam can be used to densify the exit pupil.
Labeyrie proposed the system shown in Fig.~\ref{fig: densifier} \cite{Labeyriec}
which utilises 2 micro lens arrays to re-image the exit pupils as nearly contiguous
wavefront segments. 

Here we show the first images obtained on the double star \( \alpha  \)~Gem
from a miniature hyper-telescope proving that snapshot, high dynamic range images
are obtainable from a Michelson type array.

\section{Experimental Setup}

Following the theoretical work which established the feasibility and imaging
properties of hyper-telescopes, verifications were first made through computer
simulations, followed by a laboratory system and a miniature hyper-telescope
tested on the sky. 

The latter instrument consists of a \( 100\, mm \) afocal \textbf{}refracting
telescope the aperture of which is masked with a square array of \( 8\times 8 \)
holes of \( 0.8\, mm \), regularly spaced by \( s=8\, mm \). An eyepiece produces
an exit pupil image 8 times smaller thus \textbf{}containing \( 0.1\, mm \)
sub\textbf{-}pupils spaced \( 1\, mm \) apart. The densification is achieved
with an array of micro-lenses of similar pitch located \( 100\, mm \) downstream
where the beams from each sub-pupil are spread out by diffraction in such a
way that their central lobe \textbf{}fills the facing micro-lens in the array.
These lenses having \( 100\, mm \) focal length provide parallel and nearly
adjacent collimated beams expanded from \( 0.1\, mm \) to \( 1\, mm \), achieving
a densification factor of about \( 10\, \times  \). At the focus of a lens
located immediately downstream, the central interference peak obtained is intensified
with respect to the equivalent but non densified Fizeau array. This micro-lens
array, obtained commercially, was modified by \textbf{}immersing its active
face in silicon elastomer, a medium having a slightly lower \textbf{}refractive
index than the lenses' material, in order to lower the optical path difference
(from now on OPD) caused by unequal \textbf{}thickness of the micro\textbf{-}lenses.
With its aperture diameter of \textbf{}\( D=56\, mm \) this miniature array
has \( \lambda /D=2.6^{\prime \prime } \) FWHM angular resolution (\( 1.8^{\prime \prime } \)
\textbf{}in the diagonal direction) and a usable interferometric field of view
of \( \lambda /s=18^{\prime \prime } \) for a wavelength \( \lambda =700\, nm \)
(the centre band for the detector used). Larger arrays would require adaptive
optics unless used in speckle interferometry mode. We recorded the images in
the focal plane of the interferometer array using a commercial Peltier cooled
CCD, camera with \( 9\, \mu m \) pixel size.

\section{Laboratory results}

\begin{figure}
{\par\centering \resizebox*{1\columnwidth}{!}{\includegraphics{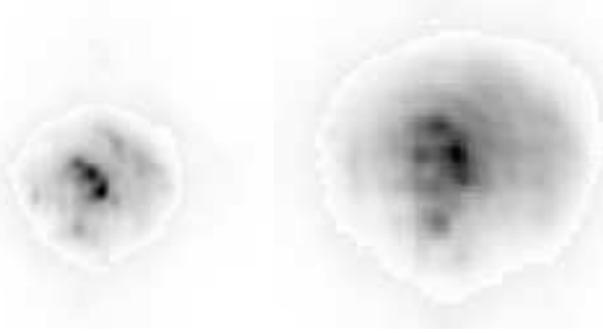}} \par}

\caption{\label{fig: lab}Laboratory images of multiple artificial stars \textbf{}obtained
with the hyper-telescope. Several pin-holes, located very close to a quartz-iodine
light bulb simulated a triple ( left) and a sextuple( right) star. Both snapshot
images were directly recorded with a CCD camera without de-convolution or image
enhancement.}
\end{figure}

The interferometer was aligned and tested in the laboratory \textbf{}with an
artificial star using a laser and a quartz-iodine bulb for white light. After
the alignment was completed the imaging capabilities of the interferometric
array were tested by replacing the pin hole placed in front of the white source
with a multi-hole mask simulating a multiple star. Fig.~\ref{fig: lab} (left)
shows a simulated triple system and a cluster of 6 objects (right). 

Owing to the unequal thickness of micro-lenses in the array, used initially
without immersion, the images obtained at this stage were not correctly phased
and exhibited a speckle pattern rather than the expected interference peak.
As expected, the speckle pattern was observed to become double when observing
a binary artificial star, thus confirming that large hyper-telescopes will be
usable for speckle interferometry when not phased adaptively.

In such large hyper-telescopes adaptive piston phasing will however be desirable,
in addition to adaptive optics within the sub-apertures. We recall that speckle
interferometry does not have an equally good imaging performance for complex
objects.

The first micro-lens array can be removed if the sub-pupil size is small enough
that diffraction suffices to fill the facing micro-lenses of the second array.
This also facilitates somewhat the phasing \textbf{}and makes the alignment
easier at the cost of an increased difficulty in controlling the size of the
diffraction lobe on the micro lenses. We discuss this effect in Sect.\( \,  \)\ref{sec: discuss}

\section{Sky results\label{sec: sky_res}}

\begin{figure}
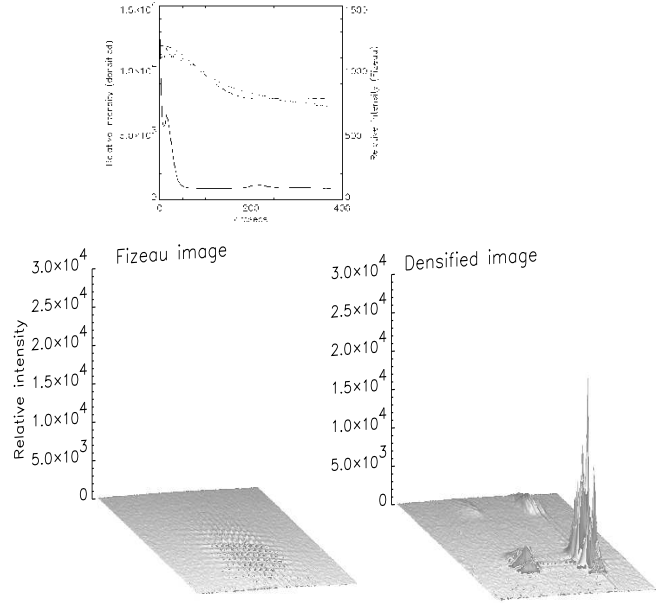

{\par\centering \resizebox*{0.8\columnwidth}{!}{\includegraphics{9868.4}} \par}

{\par\centering \resizebox*{1.15\columnwidth}{5cm}{\includegraphics{9868.5}} \par}

\caption{\label{fig: comp}Fizeau (bottom-left) and densified (bottom-right) surface
plots of the star Capella. The top plot is a \textbf{}angularly averaged intensity
profile from the bottom plots. The dashed line represents the intensity profile
for the Fizeau image, which matches the diffraction pattern of a sub--aperture,
represented by the dotted line fit. The \textbf{}solid line is the densified
intensity profile. Both snapshot images were taken in a short succession \textbf{}with
the same exposure time of \protect\( 100\, s\protect \). An intensity gain
is noticeable by \textbf{}comparing the graphs. Ghost images (left and top of
the surface plot), caused by a sub-pupil diffractive spill off of light outside
of the \textbf{}facing micro lens are visible around the densified pupil \textbf{}image.
The loss can be negligible with proper adjustments.}
\end{figure}

The miniature hyper telescope was tested on the sky (Fig.~\ref{fig: comp}\textbf{).}
This image was obtained on the star Capella (a close binary which is completely
unresolved here) by taking two separate exposures of \( 100\, s \) in the Fizeau
and densified\textbf{-}pupil mode of the hyper-telescope. \textbf{}It is of
interest to calculate the intensification achieved in the densified-pupil case
comparing the 2 images of Fig.~\ref{fig: comp}. We start off calculating the
theoretical intensity gain, in the central peak, for the 2 instrumental configurations.
\textbf{}For an unresolved star the photon count in the central peak is \( f=F_{s}(d_{0}/D_{0})^{2}, \)
where \( F_{s} \) is the photon count from a single, unresolved star, \textbf{\( d_{0} \)}
the output sub--pupil diameter and \( D_{0} \) the output pupil diameter \cite{Labeyrieb}.
In the case of our interferometer the value of \textbf{\( d_{0}/D_{0} \)} is
\( 1/70 \) with the micro-lenses removed (Fizeau) and \( 1/7 \) with the micro-lenses
in place (densified pupil). Substituting these values in the previous equation
and dividing the densified pupil photon count by the Fizeau photon count we
find a value of \( 100\times  \) for the intensification.

\begin{figure}
{\par\centering \resizebox*{1\columnwidth}{!}{\includegraphics{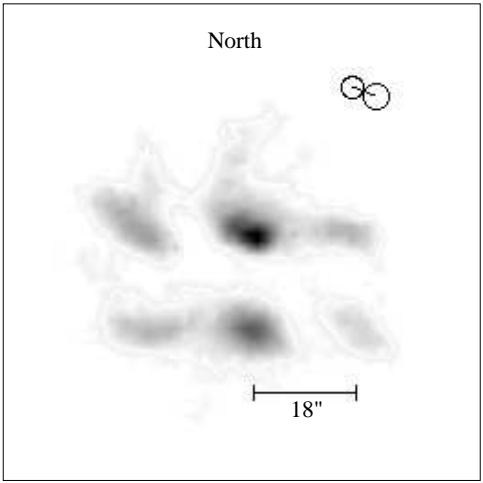}} \par}

\caption{\label{fig: double}White light image of the \textbf{}double star \protect\( \alpha \protect \)~Gem
\textbf{}produced by the miniature \textbf{}hyper-telescope. The central peak
is seen to be double in accordance to the calculated position angle and separation
of the binary star (sketched at top--right). Attenuated and dispersed \textbf{}side
lobes \textbf{}are also visible around the central peak owing to the incomplete
pupil densification and to a slight pointing offset of the telescope. The ghost
images shown in Fig.~\ref{fig: comp} are outside the picture.}
\end{figure}

We then calculate the intensification factor from the images used for generating
the plots of Fig.~\ref{fig: comp}. For the densified pupil \textbf{}image and
the \textbf{}Fizeau image, the intensity \textbf{}\( f \) was measured by integrating
the central peak. The measurements were repeated for the Fizeau and densified
configuration, using several recorded images. The comparison of the obtained
values showed an intensity gain of \( 9.2\pm 0.4\times  \) of the densified
with respect to the Fizeau configuration. When we take into account the stray
light introduced in the measurement, caused by the instrument not being properly
baffled we obtain a value of \( 24\pm 3\times  \), closer to the theoretical
value. The stray light produced a constant offset in both the Fizeau and densified
pupil image. The value of this offset was calculated from the Fizeau image by
fitting to the measured data an Airy function corresponding to the diffraction
function of the \( 0.8\: mm \) entrance apertures (Fig.~\ref{fig: comp}).
After the first minimum at \( 220^{\prime \prime } \) the function is nearly
constant and its corresponding value can be subtracted from the intensity calculation
of the Fizeau and densified pupil images. We verified that the discrepancy between
the Airy fit and the data, towards the end of the plot, was due to the detector
response threshold at low light intensity levels. This effect was found in several
other images. The data plot is also affected by photon and atmospheric noise.
The remaining discrepancy between the theoretical and the measured intensification
is probably caused by the residual optical path errors introduced by the micro-lenses
unequal thickness, the atmosphere and the misalignment of the micro-lenses with
respect to the Fizeau mask.

Given its miniature size\textbf{,} narrow field and modest collecting area (\( 41\: mm^{2} \)),
comparable to that of a naked human eye, the hyper-telescope was tested on bright
binary stars. Fig.~\ref{fig: double} shows the image obtained on \( \alpha  \)~Gem
the night of December the 3rd 1999 with a \( 30\, s \) exposure. This image
is compared to the calculated position angle and separation of the system for
\( JD\, 2451515.694 \) in Fig.~\ref{fig: double}
\begin{figure*}
{\par\centering \resizebox*{1\textwidth}{!}{\includegraphics{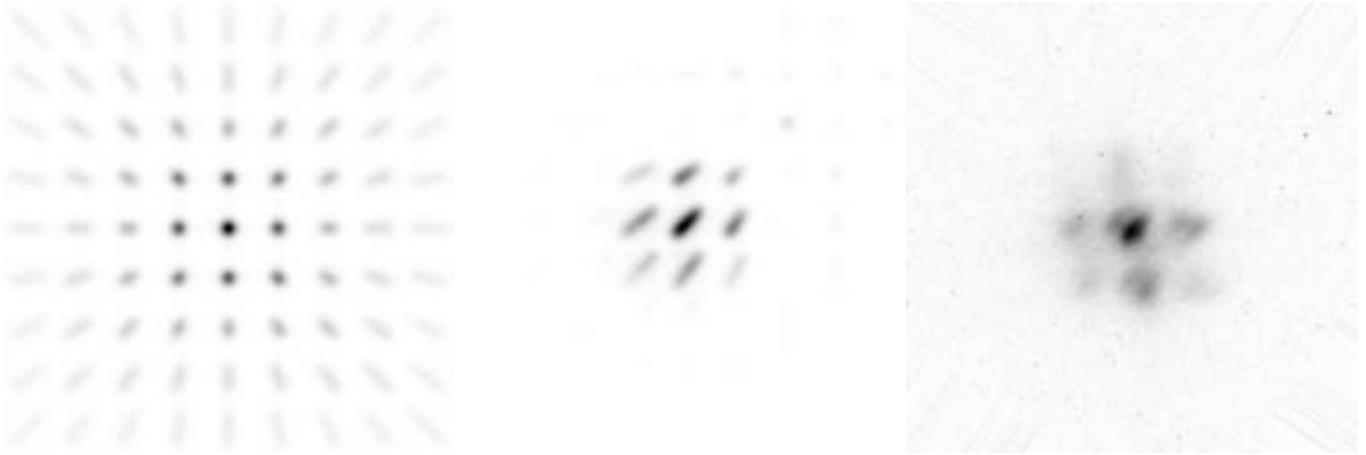}} \par}

\caption{\label{displ}Numerical simulation of PSF for the instrument (left ) with the
pupil densifier removed (Fizeau mode), in densified-pupil \textbf{}mode on a
dispersed order (centre) and the star \protect\( \alpha \protect \) Gem (right).
Comparing the star image to the numerically \textbf{}simulated PSF we deduced
that pointing errors positioned the object outside the field of view of the
interferometer. The result is a dispersed image of the double star. }
\end{figure*}

The orbital parameters of the star were obtained from the Washington Double
Star catalog. The angular separation \( \rho  \) and the position angle \( \theta  \)
were calculated using the peak pattern produced by the array. The period of
the peaks is in fact equal to \( \lambda /s \); we found a value \( \rho =4.1\pm 0.4^{\prime \prime } \)
for the angular separation. The position angle \( \theta  \) measured on the
photo centres of the double star image produced a value of \( \theta =67\pm 3^{\circ } \).
This values agree with the calculated separation \( \rho =3.97^{\prime \prime } \)
and position angle \( \theta =67.4^{\circ }. \)

\section{Discussion\label{sec: discuss}}

The use of a square grid aperture was dictated by the geometry of the micro
lenses used, which permitted to easily achieve pupil densification. Periodic
arrays, with either square or hexagonal pitch, have interference functions similarly
having a periodic array of peaks of uniform height. Non-periodic arrays, composed,
for example, of concentric rings of apertures providing little or no redundancy
\textbf{}have a speckled halo of side-lobes in their interference function.
The average level of this halo is N times fainter than the interference peak,
assumed perfectly phased.

We found that acquiring the image requires a pointing accuracy better than \( 18^{\prime \prime } \),
corresponding to the distance of the first dispersed side lobes. \textbf{}The
wavefront propagating from the star must be parallel to the main axis because
an angle of the wavefront with the axis produces an increase of the OPD among
the different apertures. As a result the white light central peak is shifted
outside the magnified diffraction function. Since the array behaves as a bi-dimensional
grating one of the dispersed peak may be found in the centre of the diffraction
function. This will still form an image but dispersed and so elongated in one
direction (Fig.~\ref{displ}). 

A disadvantage of Michelson with respect to Fizeau arrays is the limited field
of view. This is noticeable for an object moving off axis on the sky: the diffraction
envelope from the sub--pupils moves with a lower speed than the interference
peak forming the image, the speed proportional to the densification factor \cite{Labeyrieb}.
From this observation, we define the field of view for a densified pupil interferometer
as the angular extent on the sky over which the white image of stars remains
inside the diffraction envelope of the sub--pupils. Owing to the pupil densification,
the field thus defined is typically much narrower than the Airy radius of the
sub-apertures on the sky. Following this definition, \textbf{}the field of view
\textbf{}for a densified pupil interferometer \textbf{}depends on the \textbf{}amount
of densification achieved\textbf{.} Maximum densification is achieved when the
sub--pupils become adjacent\textbf{.} When this occurs the \textbf{}first dispersed
side lobes coincide with the first zero of the diffraction function. \textbf{}In
the practical case of our array \textbf{}we did not reach the maximum pupil
densification since the first side lobes are still visible. The distance between
2 apertures is \( 8\, mm \); in the image plane the angular distance between
the central white light peak and the first dispersed peak is then \( \alpha =\lambda /s, \)
giving \( \alpha  \) slightly larger than \textbf{}\( 18^{\prime \prime } \)
(\( \lambda  \) and \( s \) as previously defined). 

The basic gain of densified-pupil imaging , with respect to Fizeau, is that
the concentration of energy from a halo of many dispersed peaks into a single
white central peak intensifies the main image, thus increasing its ratio to
photon-noise. A more accurate device would contain 2 micro lens arrays working
in an afocal configuration to meet the requirements of geometrical optics as
sketched in Fig.~\ref{fig: densifier}. In the current device, shown in Fig.~\ref{fig: opt},
the diffracted Airy patterns from the sub-apertures have their feet which \textbf{}fall
upon neighbouring micro-lenses and this \textbf{}produces ``ghost images'' in
the focal plane shown in Fig.~\ref{fig: comp}. The shape of the high resolution
peak is not affected but its intensity is thus slightly reduced. Other losses
in the system are caused by the imperfect alignment of the diffracted sub-pupils
with the micro-lenses. This effect is seen as an asymmetric intensification
of the side dispersed peaks with respect to the central white light peak. Most
important perhaps, the phasing of the micro-lenses is less than perfect. This
affects the intensity of the interference peak and creates a speckled halo around
it.

\section{Conclusions}

The rather modest scale of the miniature hyper-telescope sufficed to verify
the theory of its operation and to confirm the imaging performances to be expected
with future interferometers at the kilometres scale. We have verified that arrays
in which the entrance and exit pupil are non homothetic can form direct \textbf{}images,
contrary to what was commonly believed. We have also measured an intensity gain
of \( 24\pm 3\times  \) with respect to an equivalent Fizeau array and this
is considered as supporting the \( 100\times  \) value obtainable in principle
with perfectly phased components. 

Hyper-telescope architectures are candidates for the next generation of extremely
large ground-based telescopes. Aperture sizes even larger than the 30 to 100
metres, currently considered for large steerable mosaic mirrors \cite{Gilmozzi},
can be implemented in the form of sparse mosaics. Rather than large pointing
mounts, these can extend the principle of the Arecibo radio-telescope , with
a fixed diluted primary mirror and moving focal optics, possibly carried by
stationary balloons. This configuration appears of interest even to achieve
kilometre size apertures .

In space, the situation seems \textbf{}more favourable and many sub-apertures
can be combined in formation flight to achieve kilometre size \textbf{}apertures
\textbf{}\cite{Boccaletti} \textbf{}for detecting Earth-like exo-planets \cite{Labeyriec}.

\begin{acknowledgements}
This work is based on observations with the equatorial table at the Observatoire
de Haute Provence and used hardware and software of the Observatoire de la Côte
d' Azur for data reduction. We thank the referee, Jacques Beckers, for his useful
criticism. \textbf{}One of us (Ettore Pedretti) would like to thank Farrokh
Vakili for his encouragement and many useful discussions. \bibliographystyle{aabib99}
\bibliography{9868}

\end{acknowledgements}
\end{document}